\input{aipcheck}

\documentclass[
    ,final            
  ]
  {aipproc}

\layoutstyle{6x9}

\usepackage{epsfig}

\def\kt{$k^\perp$}
\def\ktv{$\vec{k^\perp}$}
\def\ET{$E_T$}
\def\Z{$\zeta$}
\def\An{$A_N$}

\begin{document}

\title{Measurement of  Sivers Asymmetries for Di-jets in $\sqrt{s}=200$~GeV pp Collisions at STAR}


\author{J. Balewski for the STAR Collaboration}{
  address={Indiana University Cyclotron Facility, 2401 Milo B. Sampson Lane, Bloomington, IN, 47408
}
}

\begin{abstract}
Measurement of the transverse spin dependence of the di-jet opening angle in pp collisions at $\sqrt{s}=200$~GeV has been performed by the STAR collaboration. An analyzing power consistent with zero has been observed over a broad range in pseudorapidity sum of the two jets with respect to the polarized beam direction. A non-zero (Sivers) correlation between transverse momentum direction of partons in the initial state and transverse spin orientation of the parent proton has been previously observed in semi-inclusive deep inelastic scattering (SIDIS).
 The present measurements are much smaller than deduced  from predictions made for STAR di-jets based on non-zero quark Sivers  functions deduced from SIDIS, and furthermore indicate that gluon Sivers asymmetries are comparably small.
\end{abstract}

\maketitle



Experimental determination of the contribution from orbital angular momentum of the proton's constituents to its spin is  difficult. A possible manifestation of orbital momentum are Sivers correlations~\cite{Siv90} between the transverse spin ($\vec{S_T}$) direction of a polarized proton and transverse momentum (\ktv) directional preferences of unpolarized partons carrying longitudinal momentum fraction $x$ of the proton:
\begin{equation}
\frac{1}{2}\Delta^Nf(x,k^\perp)
 \frac {\vec{P} \cdot \left( \vec{k^\perp } \times~  \vec{S_T}\right)}
        {|\vec{P}|\cdot| \vec{k^\perp}| \cdot |\vec{S_T}| }
\end{equation}
where $\vec{P}$ is the momentum vector of the polarized proton and ~$\Delta^Nf(x,k^\perp)$ is called the Sivers function.

Recently measurement of a non-zero Sivers function for $\pi^+$ in SIDIS has been reported by the HERMES collaboration \cite{Her05}. The same group found an asymmetry consistent with zero  for $\pi^-$, leading to an interpretation~\cite{Vog05} that the  Sivers functions are opposite in sign and different in magnitude for $u$ vs. $d$ quarks.
Boer and Vogelsang~\cite{Boe04} suggested that the Sivers effect would be manifested as a spin-dependent azimuthal side preference for almost back-to-back di-jets produced in proton collisions involving transversely polarized beam.

In this paper we  report Sivers single spin asymmetry (\An) measurements for di-jet production in  kinematic regions that allow us to distinguish quark from gluon Sivers functions. We compare our results to  predictions~\cite{Vog05} derived from SIDIS, in order to judge the universality of the extracted quark Sivers functions.

\begin{figure}
\centering \leavevmode \epsfverbosetrue \epsfclipon \epsfxsize=12.cm\epsffile[0 240 570 470 ]{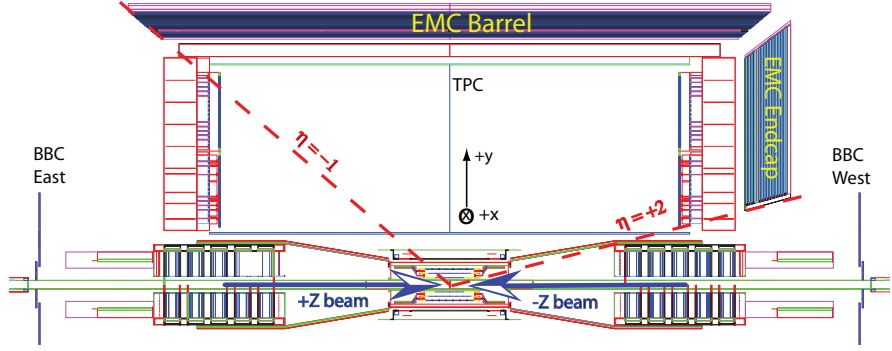}

\label{starGeo}
  \caption{Geometry of the STAR detector. }
\end{figure}


The di-jet measurement has been performed in the 2006 run using the STAR detector at RHIC \cite{Ack03}. A 3 week-long run resulted in a sampled luminosity of 1.1~pb$^{-1}$.
Two polarized beams collided in the center of the STAR detector, as shown in Fig.~\ref{starGeo}. Electromagnetic calorimeters (EMC) covering pseudorapidity range $\eta \in [-1,+2]$ and full azimuthal angle $\phi$ have been used to trigger on di-jets. 
A dedicated software trigger accepted events with two EMC patches covering $0.6 \times 0.6$ in $\Delta\eta \times \Delta\phi$,  separated azimuthally by at least 60$^\circ$, and  containing at least 3.3 and 3.6~GeV of transverse energy (\ET), respectively. Additionally, the hardware trigger required  $E_T\ge14$~GeV  summed over  the whole EMC and a coincidence between 2 forward segmented scintillators (BBC) placed at  $3.3\le|\eta| \le 5.0$ on either side of the interaction point.
In this analysis we have used the on-line values of $\eta$ and $\phi$ for both jets obtained from $E_T$-weighted centroids of the EMC tower locations included in the jet patches.
The results reported here include only those events where both detected jets had  EMC $E_T\le8.0$~GeV within their respective  $0.6 \times 0.6$  $\Delta\eta \times \Delta\phi$ patch.

\begin{figure}
\centering \leavevmode \epsfverbosetrue \epsfclipon \epsfxsize=15.cm\epsffile[10 430 570 600  ]{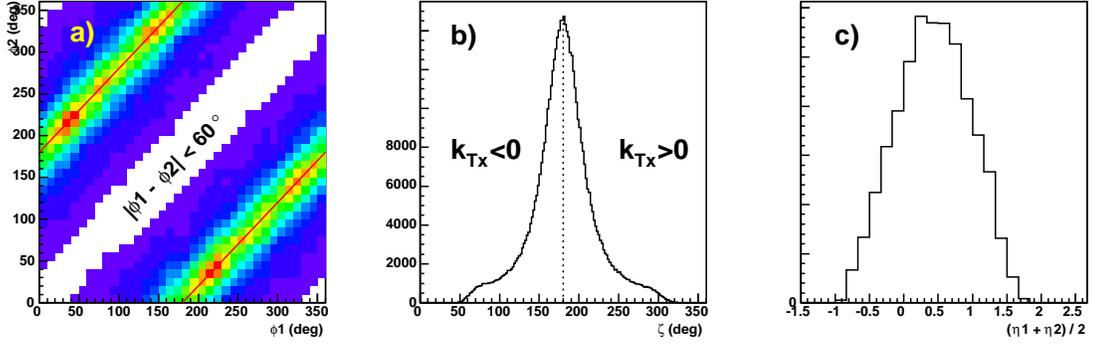}
  \caption{Characteristic distribution of events that pass the STAR di-jet trigger:  a) azimuthal correlation, b) signed opening angle \Z, and c) averaged pseudorapidity, all from trigger-level EMC information only.}   
  \label{diJetYield}
\end{figure}

 The azimuthal  correlation between the two jets is shown in Fig.~\ref{diJetYield}a,
 where back-to-back di-jets dominate the spectrum. For the Sivers analysis, we define the ``signed'' azimuthal di-jet opening angle \Z~ (Fig.~\ref{diJetYield}b), chosen to be $>180^\circ$ when $cos\phi_b>0$ and $<180^\circ$ otherwise, where $\phi_b$ is the di-jet bisector angle measured with respect to STAR's $+x$-axis (see Fig.~\ref{starGeo}). The spectrum of average  $\frac{\eta_1+\eta_2}{2}$ is shown in Fig.~\ref{diJetYield}c.
The most forward measured di-jets when $\eta_1+\eta_2>2$ involve preferentially  quark (gluon) from the polarized beam heading toward +z (-z) direction.


The  vertical beam spin orientation alternated for each successive bunch of one beam and each pair of bunches of the other.
An integrated Sivers \An~ was extracted for each beam from spin-sorted \Z-spectra using a cross-ratio in which ``left'' and ``right'' yields were deduced from $\zeta >180^\circ$~and~$\zeta<~180^\circ$, respectively, for the $+z$ beam and the opposite assignment was used for the $-z$ beam.
 On-line values of beam polarization have been  used and a systematic normalization error of $\pm20$\% is assigned to \An~ until final beam polarization values are known. The cross-ratio method does not require independent luminosity monitors if the detector has approximately equal left-right efficiency, as is true for this measurement.

The jet direction  deduced from its electromagnetic component only can deviate from the direction of the parent parton due to fragmentation processes and finite detector resolution. Simulations with the Pythia 6.205 event generator and full GEANT modeling of the STAR detector have been used to estimate this azimuthal smearing to be $\sigma\approx6.3^\circ$. This $\phi$-smearing is small compared to the observed overall \Z~ width of $\sim 20^\circ$, driven by the intrinsic \kt~ smearing of the partons inside the proton.  However, this $\phi$-smearing  does cause a dilution of the measured \An~ by  10\% of the input parton-level asymmetry, as determined from a toy model simulation.

The toy model assumed a binary parton collision with an  initial Gaussian \ktv~ distribution  with added  exponential tails. To simulate the Sivers effect it assumed a small offset in the centroid of the model  \ktv~ distribution, in the direction $\vec{S_T} \times \vec{P}$, using the randomly assigned spin direction for each beam.
 Azimuthal directions of the outgoing partons were smeared with a Gaussian of $\sigma=6.3^\circ$ to mimic fragmentation and detection processes. The toy model parameters were tuned to reproduce the observed jet $E_T$- (from full jet reconstruction including charge-particle tracks) and $\zeta$-spectra.
  
\begin{figure}
\centering \leavevmode \epsfverbosetrue \epsfclipon \epsfxsize=11.cm \epsffile[10 370 570 700  ]{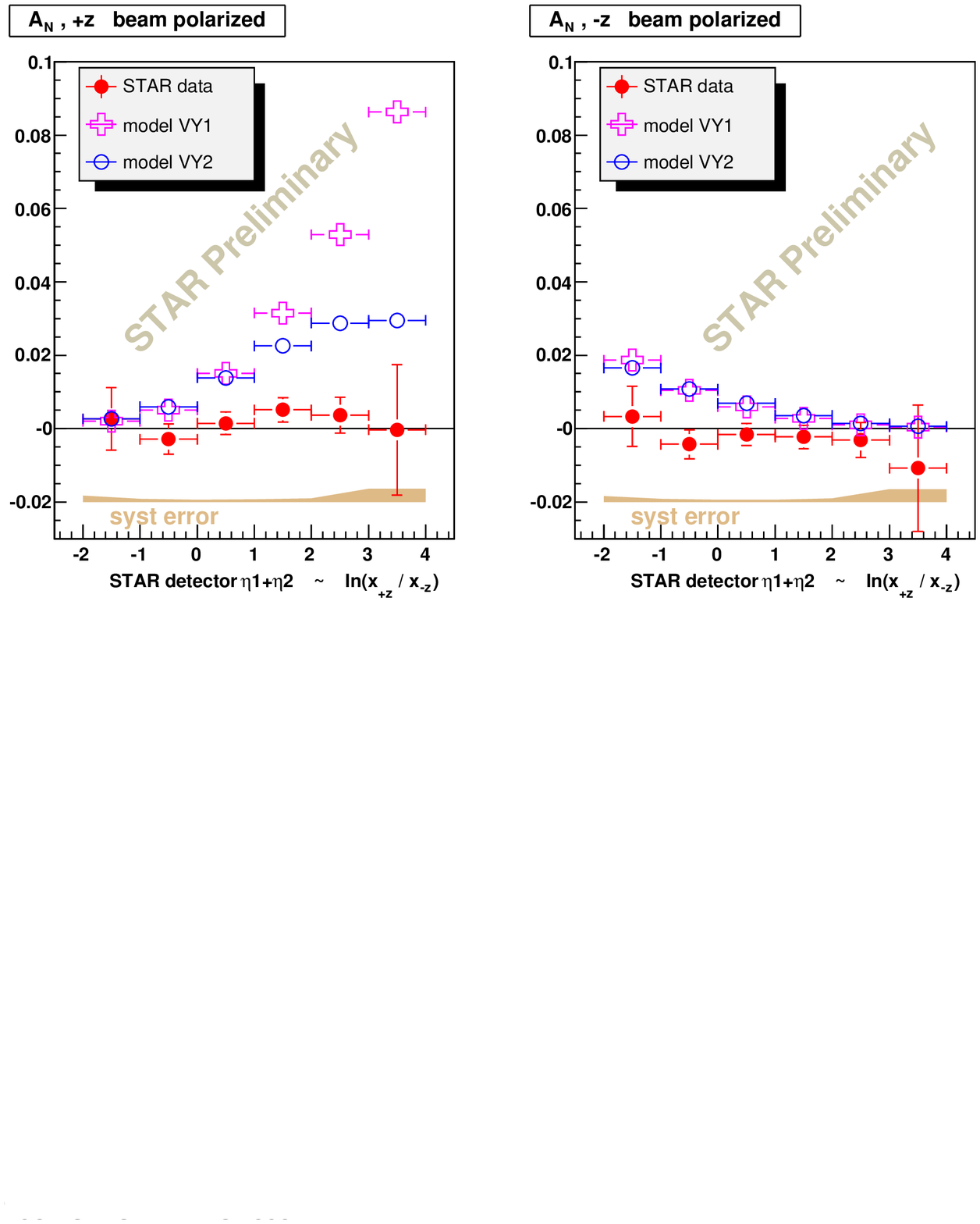}
  \caption{Comparison of STAR results for $A_N^{+z}$ (left) and  $A_N^{-z}$  (right) as function of $\eta_1+\eta_2$ with predictions \cite{Vog05}, based on two models of quark Sivers functions deduced from fits to HERMES SIDS results \cite{Her05}.}
  \label{SivAn}
\end{figure}

Fig.~\ref{SivAn} shows the measured analyzing power for both beams (heading toward $\pm z$)  compared to predictions \cite{Vog05} as function of $\eta_1+\eta_2$ integrated over the full  STAR EMC $\eta$ acceptance and for $|\zeta -180^\circ | \le68^\circ$. The full analyzed sample yields $2.6\times 10^6$ di-jets with average $A_N^{\pm z}$ values consistent with zero for both beams, within statistical uncertainties of $\pm 0.002$. Based on the toy model these results probe  a Sivers $\left< k^x_T \right>$ preference as small as $\pm 3$~MeV, or $\pm 0.2$~\% of the inferred rms width of the $k^{x,y}_T$ distributions.

The systematic error bands in Fig.~\ref{SivAn} arise from two sources combined in quadrature. An error of $\pm20$\% of the statistical uncertainty on each bin allows for multi-jet event contribution to the \Z~ distribution wings (see Fig.~\ref{diJetYield}b). In addition, we include a model-dependent uncertainty on the dilution factor ($0.90\pm0.05$) deduced from the toy model. The 20\% uncertainty on the on-line beam polarization values is not included in the error band.

The predictions \cite{Vog05} in Fig.~\ref{SivAn} are based on two models (labeled VY1 and VY2) of the
$u$ and $d$ quark Sivers functions fitted to HERMES SIDIS results \cite{Her05}.
 The models
differ in tying the d-quark Sivers function shape to either the
$u(x)$ (VY1) or $d(x)$ (VY2) unpolarized parton distribution function. Both calculations assume zero gluon
Sivers function. The predictions are integrated over a jet
$p_T$ range (5-10 GeV/c) well matched to our measurements, and further over the STAR detector acceptance
within each $\eta_1+\eta_2$ bin.
In these calculations,
  di-jet production was treated like Drell-Yan di-lepton production, with gauge link factors for initial-state interactions only~\cite{Vog05}.
 We have reversed the sign of the predictions to adhere to the Madison convention used for our measured asymmetries.

In the region $\eta_1+\eta_2 \ge 2$, $A_N^{+z}~ (A_N^{-z})$ preferentially probes quark (gluon) Sivers functions. The
 predicted  $A_N^{+z}$ values in this region reflect the sizable HERMES Sivers
asymmetries, while  $A_N^{-z} \simeq 0$ because  gluon Sivers effects have been ignored.

 All measured di-jet asymmetries are within two standard deviations of zero, suggesting that both quark and gluon Sivers effects are considerably smaller for di-jet production in pp collisions than are the quark Sivers effects inferred from the HERMES results.

The experiment-theory discrepancies in Fig.~\ref{SivAn} might reflect a breakdown of pQCD factorization for the soft \kt~ values probed for back-to-back di-jets (near $\zeta=180^\circ$).
 It remains an open challenge for theory to account simultaneously for the present di-jet
results and for the sizable transverse single-spin asymmetries seen in SIDIS \cite{Her05} and in forward $pp \rightarrow \pi^\circ X$ \cite{Ada04}.

\bibliographystyle{aipprocl} 


\end{document}